\documentclass[twoside,11pt]{article}

\usepackage{color}
\usepackage{listings}

\usepackage{jmlr2e}

\newcommand{\sVec}{\vec{s}}
\newcommand{\yVec}{\vec{y}}

\newcommand{\disT}{\textstyle}

\newcommand{\Prosper}{{\tt ProSper}}

\ShortHeadings{ProSper}{Exarchakis et al.}
\firstpageno{1}

\definecolor{Red}{rgb}{1.0,0,0} 

\definecolor{MyGreen}{rgb}{0,0.5,0}

\begin{document}


\title{ProSper - A Python Library for Probabilistic Sparse Coding with Non-Standard Priors and Superpositions}

\author{\name Georgios Exarchakis$^*$ \email georgios.exarchakis@inserm.fr \\
       \addr Sorbonne Universit\'e, INSERM, CNRS, Institut de la Vision, Paris, France
       \AND 
       \name J\"org Bornschein$^{*,\dagger{}}$ \email j.bornschein@gmail.com \\
       \addr FIAS, Goethe-Universit\"at Frankfurt am Main, Germany
       \AND 
       \name Abdul-Saboor Sheikh \email sheikh.abdulsaboor@gmail.com \\
       \addr Zalando Research, Berlin, Germany
       \AND 
       \name Zhenwen Dai \email zdai@sheffield.ac.uk \\
       \addr Dept of Computer Science, University of Sheffield, UK
       \AND 
       \name Marc Henniges \email marc.henniges@d-fine.com\\
       \addr d-fine GmbH, Frankfurt, Germany
       \AND
       \name Jakob Drefs \email jakob.drefs@uol.de\\
       \addr Machine Learning Division, C.\,v.\,Ossietzky Universit\"at Oldenburg, Germany
       \AND
       \name J\"org L\"ucke \email joerg.luecke@uol.de \\
       \addr Machine Learning Division, C.\,v.\,Ossietzky Universit\"at Oldenburg, Germany
}
\editor{Unknown}

\maketitle
\ \\[-10mm]
\small\hspace{-4ex}\,$^*$contributed equally, $^{\dagger}$now at DeepMind, London, UK\\[3mm]
\normalsize

\begin{abstract}
{\tt ProSper} is a python library containing probabilistic algorithms to learn
dictionaries. Given a set of data points, the implemented algorithms seek to
learn the elementary components that have generated the data. The library widens
the scope of dictionary learning approaches beyond implementations of
standard approaches such as ICA, NMF or standard $L_1$ sparse coding. The
implemented algorithms are especially well-suited in cases when data
consist of components that combine non-linearly and/or for data requiring
flexible prior distributions. Furthermore, the implemented algorithms go beyond
standard approaches by inferring prior and noise parameters of the data, and
they provide rich a-posteriori approximations for inference. The library is
designed to be extendable and it currently includes: Binary Sparse Coding
(BSC), Ternary Sparse Coding (TSC), Discrete Sparse Coding (DSC), Maximal Causes Analysis (MCA), Maximum
Magnitude Causes Analysis (MMCA), and Gaussian Sparse Coding (GSC, a recent
spike-and-slab sparse coding approach). The algorithms are scalable due to a combination of variational
approximations and parallelization. Implementations of all algorithms allow for
parallel execution on multiple CPUs and multiple machines for medium to
large-scale applications. Typical large-scale runs of the algorithms can use 
hundreds of CPUs to learn hundreds of dictionary elements from data with
tens of millions of floating-point numbers such that models with 
several hundred thousand parameters can be optimized.
The library is designed to have minimal dependencies and to be easy to use. 
It targets users of dictionary learning algorithms and Machine
Learning researchers. 
\end{abstract}
%
%
\section{Introduction}
Dictionary learning is a broad subfield of Machine Learning with numerous applications in different data domains.
It addresses the unsupervised extraction of latent components or factors of observed data samples. 
Classical examples of dictionary learning methods include deterministic approaches such as K-SVD, ICA, projection pursuit, NMF among many others.
In contrast to deterministic approaches, probabilistic methodologies for dictionary learning are based on a generative data model
to yield a probabilistic objective (typically the data likelihood) for optimization.
While some probabilistic approaches such as sparse coding with a Gaussian noise model and Laplace prior \citep{OlshausenField1996} closely link to popular deterministic $L_1$-regularized sparse coding, many other choices of priors do not have a corresponding counterpart. Similarly, it is straight-forward to define non-standard probabilistic data models, e.g., by choosing the component superposition assumption to be different from linear, which can be a more reasonable choice for many types of data \citep{BornscheinEtAl2013,DaiEtAl2013,SheikhEtAl2019}. In several contributions, it was shown over the past years that such non-standard sparse coding models can efficiently be trained at large scales and with large data sets. Furthermore, parameters other than basis functions (i.e., dictionary elements) can be learned using recent approximate learning methods, notably the sparsity level and the data noise. The \Prosper{} library contains such novel and non-standard sparse coding algorithms in a unified python software framework. Most notably, the used priors are binary, ternary, categorical or follow a spike-and-slab distribution, and the superposition models of components are linear or non-linear. Based on truncated posterior approximations \citep{LuckeEggert2010} and MPI parallelization for many CPU nodes and cores,
all algorithms can be efficiently applied to large data sets and large dictionary sizes 
\citep{HennigesEtAl2010,GuiraudEtAl2018,ExarchakisEtAl2012,ExarchakisLuecke2017,LuckeSheikh2012,SheikhEtAl2014,PuertasEtAl2010,SheikhEtAl2019,BornscheinEtAl2013}.
The software uses the {\tt NumPy} and {\tt SciPy} packages to define data and parameter containers
and apply elementary numerical operations. \Prosper{} algorithms are also enabled for parallel 
computing using the {\tt MPI for Python} package and a data logging utility built on top of the
{\tt PyTables} package.

%
%





\newcommand\sbreak{\\[1mm]}

 \begin{table*}[t]
 \caption{
 List of algorithms with their superposition models for component combination and their assumed distributions for observed and hidden variables.\vspace{-3mm}
}
 \begin{center}
 \small
 \renewcommand{\arraystretch}{1.1}
 \newcommand{\myhspace}{\hspace{-1mm}}
 \begin{center}
 \begin{tabular}{|c|ccccc|}\hline
  {\bf Model}          & \multicolumn{5}{c|}{\phantom{$\int^A$}{\bf Properties}\phantom{$\int^A$}} \\[1mm]\hline
       Acronym                    &          \myhspace{}superposition\myhspace{}              & \myhspace{}obs.\ variables\myhspace{} & \myhspace{}noise type\myhspace{}   &\myhspace{}hidden variables\myhspace{} &\myhspace{} prior model\myhspace{} \\[1mm]\hline
         BSC   &   linear   & real &   Gaussian  &  binary \{0,1\} &    Bernoulli\\
         TSC   &   linear   & real &   Gaussian  &  ternary \{-1,0,1\} &  categorical/zero-mean\\
         DSC   &   linear   & real &   Gaussian  &   discrete &  categorical\\
         GSC   &   linear   & real &   Gaussian  &  real &  spike-and-slab\\
         MCA   &   $\max{}$ & real &   Gaussian  &  binary \{0,1\} &    Bernoulli\\
        MMCA   &   $|\max{}|$ & real &   Gaussian  &  binary \{0,1\} &    Bernoulli\\
         GMM   &   none    & real &   Gaussian  &  one integer &  categorical\\
         PMM   &   none    & integer ($\geq{}$0)  &   Poisson  &  one integer &  categorical \\\hline
 \end{tabular}
 \end{center}
 \label{TabDetails}
 \end{center}
 \end{table*}
 \begin{table*}[h]
\ \\[-15mm]
 \caption{
 List of Prosper models together with their associated references (main ref.\ bold).\vspace{-2mm}
}
 \begin{center}
 \footnotesize
 \renewcommand{\arraystretch}{1.1}
 \newcommand{\myhspace}{\hspace{-1mm}}
 \begin{center}
 \begin{tabular}{|c|c|c|}\hline
    Model  &   Full Name &  References (bold for main reference) \\[1mm]\hline
   BSC         & Binary Sparse Coding &             L\"ucke \& Eggert {\em JMLR} 2010\\
               &              &  {\bf Henniges et al., {\em Proc. LVA/ICA} 2010} \nocite{HennigesEtAl2010}\\
               &              &  Guiraud et al., {\em GECCO}, 2018               \nocite{GuiraudEtAl2018}\sbreak
   TSC         & Ternary Sparse Coding  & {\bf Exarchakis et al., {\em Proc. LVA/ICA} 2012} \nocite{ExarchakisEtAl2012} \sbreak
   DSC         & Discrete Sparse Coding  & {\bf Exarchakis et al., {\em Neural Comp.} 2017} \nocite{ExarchakisLuecke2017}\sbreak
   GSC         & Gaussian Sparse Coding  &  L\"ucke \& Sheikh, {\em Proc. LVA/ICA} 2012 \nocite{LuckeSheikh2012}\\  
               & (spike + Gaussian slab) &  {\bf Sheikh et al., {\em JMLR} 2014}        \nocite{SheikhEtAl2014}\sbreak
   MCA         & Maximal Causes Analysis & L\"ucke \& Sahani, {\em JMLR} 2008 \nocite{LuckeSahani2008}\\
               &   &  L\"ucke \& Eggert {\em JMLR} 2010                       \nocite{LuckeEggert2010} \\
               &              &  {\bf Puertas et al., {\em NIPS} 2010}        \nocite{PuertasEtAl2010}\\
               &              &  Sheikh et al., {\em PLOS Comp.\ Bio.} 2019.  \nocite{SheikhEtAl2019}\sbreak
   MMCA         & Max Magnitude MCA  & {\bf Bornschein et al., {\em PLOS Comp.\ Bio.}, 2013} \nocite{BornscheinEtAl2013}\sbreak
   GMM         & Gaussian Mixture Model & standard EM for GMM algorithm\sbreak
   PMM         & Poisson Mixture Model  & standard EM for a Poisson mixture\\\hline
 \end{tabular}
 \end{center}
 \label{TabProtocols}
 \end{center}
\vspace{-5mm}
 \end{table*}

\section{Learning Algorithm and Data Models}
All the models included in {\tt ProSper} are based on the following data generative process:
%
%
\begin{eqnarray}
\disT{}p(\sVec\,|\,\Theta) &=& \textrm{latent variable prior distribution, e.g. Bernoulli for BSC}\label{EqnBSCPrior} \\
\disT{}p(\yVec\,|\,\sVec, \Theta) &=& \disT{}p(\vec{y};\,\vec{f}(\Theta, \vec{s}))\ \ \ \  (\textrm{noise model})\label{EqnBSCNoise},
\end{eqnarray}
where $\Theta$ is the set of model parameters (typically containing the dictionary $W$ but also prior and noise parameters). The models can be fully specified by defining Eq.\,\ref{EqnBSCPrior} and \ref{EqnBSCNoise}, i.e., they are categorized based on the noise distribution, the prior distribution, and the function $\vec{f}$ that
determines the influence of the latent variables on the observed variables ($\vec{f}$ can be thought of as a link function). The most common instance of the function $\vec{f}$ for dictionary learning is a linear function, e.g.\ $\vec{f}(\Theta, \vec{s})=W\vec{s}$, but the library allows for alternative choices.
Tab.\,\ref{TabDetails} lists the data models of the currently implemented algorithms.

All algorithms use expectation maximization for parameter optimization and truncated posteriors as efficient approximation \citep{LuckeEggert2010}.
The approximation method has been successfully applied in numerous contexts to the models listed in Tab.\,\ref{TabDetails}. A list of scientific publications describing the models 
along with their specific implementation details for inference and learning can be found in Tab.\,\ref{TabProtocols}.

\noindent

\section{User Interface and Documentation}
The interface is designed to be as reusable and flexible as possible. We use three objects that compose a learning algorithm: {\tt Annealing}, {\tt Model}, and {\tt EM}. The learned parameters are contained in a python dictionary that is shared among these objects. The {\tt Annealing} object is responsible for the schedule of the learning algorithm and defines methods {\tt next}, and {\tt reset} to specify the step and technical interventions of the training process. The {\tt Model} object, probably the most crucial element of our library, defines methods {\tt step} and {\tt standard\_init} which respectively define one optimization step of the algorithm, and the initialization of the parameters. The {\tt run} method of the {\tt EM} object combines {\tt Annealing} and {\tt Model} by running a loop over the optimization step of the model modified as specified in the {\tt Annealing} object.

Annealing schedules are specified by objects that inherit from the abstract {\tt Annealing} class. As an example, we provide the class {\tt LinearAnnealing} that controls the number of iterations of the training algorithm, parameter noise, and deterministic annealing of the approximate posterior. 
Similarly, {\tt Model} instances inherit from the abstract {\tt Model} class and implement the relevant methods. We provide implementations for the models listed in Tab.~\ref{TabProtocols}. 

To run an algorithm, we start by instantiating a model with corresponding hyperparameters, e.g.\ {\tt model = BSC\_ET(D, H,  Hprime, gamma)} were {\tt D} and {\tt H} are observed and latent dimensions respectively and {\tt Hprime} and {\tt gamma} approximation parameters. We proceed by initializing the annealing class, e.g.\ {\tt anneal = LinearAnnealing(150)}. Using a dictionary to store the data under the key {\tt `y'} we call the {\tt standard\_init} method to randomly initialize the parameters, e.g. {\tt params = model.standard\_init(\{"y":data\})}. The {\tt EM} class is initialized as {\tt em = EM(model=model, anneal=anneal)} and we train the model with {\tt em.run()}. 
We can then apply the {\tt inference} method to extract approximate posterior information about the data, e.g. {\tt res=model.inference(anneal,em.lparams,data)}. This yields, e.g. the (estimated) most likely latent variable configurations ({\tt res["s"]}) and corresponding approximate posterior probabilities ({\tt res["p"]}) for each data point as well as additional information specific for each model.
\section{Related Software Libraries}
Most libraries for sparse coding or dictionary learning are based on deterministic objectives:
The SPAMS library \citep{MairalEtAl2010,MairalBPS09} contains a collection of deterministic sparse coding algorithms with 
$L_1$, $L_2$ and $L_{\infty}$ regularization (C++ based, interfaces to Matlab, R and Python).
Similarly {\tt MLPack} \citep{mlpack2018} contains standard $L_1$/$L_2$ regularized sparse coding. Scikit-learn \citep{scikit-learn} also
contains a standard (deterministic and $L_1$ regularized) sparse coding version and
provides standard NMF or LDA implementations. 
{\tt Sparsenet} and {\tt glm-ie} both use a continuous and probabilistic sparse coding data model, and both require the data model to provide
monomodal a-posterior distributions for convex optimization. The data models used by {\tt Sparsenet} or {\tt glm-ie} consequently do not overlap
nor do the parameter optimization methods. Sparsenet implements the original algorithm by \citet{OlshausenField1996} based on MAP, and {\tt glm-ie} \citep{glm-ie2012} provides sophisticated inference for the generalized (sparse) linear model. 
{\tt libDAI} \citep{Moij2010} and {\tt Libra} \citep{lowd&rooshenas2015} are libraries for inference and learning in general graphical models. 
None of the libraries is as optimized for probabilistic sparse coding as {\tt ProSper} or provide its efficient variational EM approach. But {\tt libDAI} and {\tt Libra} are much more general in the graphical data models that can be treated. While {\tt libDAI} focuses more on inference, {\tt Libra} focuses more on learning and structure learning.


\acks{We would like to thank Julian Eggert for the many discussions. We would also like to acknowledge support by the DFG in grants 
SFB 1330 (HAPPAA, ID 352015383) and EXC 2177/1 (Hearing4A, ID 390895286). Early stages of the work were funded by the DFG (grant LU 1196/4-2) and the BMBF (grant 01GQ0840).}

\bibliography{cnml-all}

\end{document}